\documentclass[aps,prb,twocolumn,superscriptaddress]{revtex4-1}

\usepackage{graphicx}
\usepackage{amsmath}
\usepackage{subfigure}
\usepackage{color}

\begin{document}

\title{ Unconventional Large Linear Magnetoresistance in Cu$_{2-x}$Te}

\author{Ali A. Sirusi}
\affiliation{Department of Physics and Astronomy, Texas A\&M University, College Station, Texas 77843, USA}
\author{Alexander Page}
\affiliation{Department of Physics, University of Michigan, Ann Arbor, MI 48109, USA}
\author{Lucia Steinke}
\affiliation{Department of Physics and Astronomy, Texas A\&M University, College Station, Texas 77843, USA}
\author{Meigan C. Aronson}
\affiliation{Department of Physics and Astronomy, Texas A\&M University, College Station, Texas 77843, USA}
\affiliation{Department of Materials Science and Engineering, Texas A\&M University, College Station, Texas 77843, USA}
\author{Ctirad Uher}
\affiliation{Department of Physics, University of Michigan, Ann Arbor, MI 48109, USA}
\author{Joseph H. Ross, Jr.}
\affiliation{Department of Physics and Astronomy, Texas A\&M University, College Station, Texas 77843, USA}
\affiliation{Department of Materials Science and Engineering, Texas A\&M University, College Station, Texas 77843, USA}

\date{\today}
\begin{abstract}
We report a large linear magnetoresistance in Cu$_{2-x}$Te, reaching 
$\Delta\rho/\rho(0)$ = 250\% at 2 K in a 9 T field.
This is observed for samples with $x$ in the range 0.13 to 0.22, and the results are
comparable to the effects observed in Ag$_2 X$ materials, although in this case the results appear for a much wider
range of bulk carrier density. Examining the magnitude vs. crossover field from low-field quadratic
to high-field linear behavior, we show that models based on classical transport behavior best explain the observed
results. The effects are traced to misdirected currents due to topologically inverted behavior in this system,
such that stable surface states provide the high mobility transport channels.
The resistivity also crosses over to a $T^2$ dependence in the temperature range where the large linear MR appears, 
an indicator of electron-electron interaction effects within the surface states. Thus this is an example of a system in which 
these interactions dominate the low-temperature behavior of the surface states.
\end{abstract}

\maketitle

Considerable attention has recently been devoted to systems exhibiting linear magnetoresistance (MR),
starting with the effects observed in Ag$_2$Te and Ag$_2$Se \cite{Xu97,Husmann2002}. In these
systems linear MR extends to very large applied fields, in contrast to 
conventional conductors exhibiting quadratic MR which eventually saturates with increasing field.
The origin of this effect can be traced to high mobility topological surface states in these 
systems \cite{Zhang2011,Lee2012,sulaev13,Kim2016},
and meanwhile many related systems have been discovered to exhibit such behavior, including
Dirac semimetals \cite{Novak2015,Narayanan2015,Song2015}.

Models proposed to explain the Ag$_2 X$ effects include the quantum
mechanism of Abrikosov \cite{Abrikosov1998}, with low mass carriers
generated through disorder-induced band contacts leading to orbital quantization 
in relatively small fields. Weak anti-localization \cite{Hikami1980,veldhorst2013} can also be important for 
magneto-transport of topological surface states, and the interplay of these effects with linear MR has been 
examined, for example, in Bi$_2$Te$_3$-based topological insulators \cite{Assaf2013,Tian2014}. 
On the other hand 
Parish and Littlewood \cite{Parish2003,Parish2005} showed that a classical mechanism will give 
linear MR extending over a wide range of fields in the case of a distribution
of carrier mobilities significantly exceeding the mean value ($\Delta\mu \gg \langle\mu\rangle$).
Herring\cite{Herring1960} had earlier shown that linear MR may occur in weakly inhomogeneous systems,
for fields where the cyclotron orbit period exceeds the scattering time, equivalent to $B\langle\mu\rangle > 1$. 
A common feature of these models is low mass/high mobility states, although 2 dimensional surface 
states are not specifically required.

With the Ag$_2$Se linear MR observed to track with
mobility \cite{vonkreutzbruck09}, a classical model would appear to apply. Nevertheless the origin remains 
unclear because of the small fields needed to limit massless surface states to the lowest Landau 
level \cite{Zhang2011}, thus suggesting a quantum origin. It has further been 
proposed \cite{Schnyders2015} that more conventional processes involving compensating charge carriers may 
dominate in Ag$_2$Te, and a mechanism based on spin splitting of surface states has also been
advanced for topological insulators \cite{Wang2012b}. Weak antilocalization as
a bulk rather than surface effect \cite{veldhorst2013} may also occur in these layered systems in the presence of spin-orbit coupling.
Furthermore it was recently demonstrated \cite{Song2015} that even very weak disorder 
may lead to such effects in 3D high-mobility systems such as the Dirac semimetals. 

Cu$_2$Te has been of significant interest for potential applications including
thermoelectric and solar energy conversion, as well as a variety of 
nano-devices \cite{ballikaya13,He2015,nguyen13,poulose16}, and it has
been connected to a topologically nontrivial band configuration \cite{Sirusi2016,Ma2016}. 
Synthesized materials in bulk have a Cu$_{2-x}$Te stoichiometry, with the Cu deficit corresponding to vacancies 
which lead to $p$-type semiconducting behavior. 
Here we present magnetotransport properties of materials in the range $x$ = 0.13 to 0.22,
exhibiting a large linear MR which 
can be traced to surface states, reinforcing the topological insulator nature of this system, and occurring in a regime 
of high carrier density and with strong electron interactions 
distinct from what has been observed in other systems.


The three polycrystalline Cu$_{2-x}$Te samples were obtained by solid state reaction and vacuum annealing.
Their properties have been described in more 
detail in Ref. \onlinecite{Sirusi2017}. Compositions from electron microprobe measurements are 
Cu$_{1.87}$Te, Cu$_{1.82}$Te, and Cu$_{1.78}$Te ($x$ = 0.13 to 0.22), with
Hall measurements showing them to be heavily-doped $p$-type semiconductors with room 
temperature carrier densities 3.6, 4.1, and 6.5 $\times 10^{21}$ cm$^{-3}$, respectively.
The results along with NMR measurements are consistent with a Fermi level in the bulk which is pulled below the 
valence band edge due to Cu deficit \cite{Sirusi2017},
with room temperature Hall results matching the expected bulk carrier densities.
The structure for Cu$_{1.87}$Te and Cu$_{1.82}$Te is a superstructure of the hexagonal Nowotny 
structure \cite{Nowotny,Sirusi2017}, with a somewhat different superstructure
for the Cu$_{1.78}$Te case. 
Measurements reported here utilized a Quantum Design PPMS system and a Quantum Design MPMS 
combined with an AC bridge. 
Transport measurements were performed on bar-shaped samples cut from the polycrystalline ingots,
with magnetoresistance measured in the geometry with the field perpendicular to the current direction.

\begin{figure}		
\includegraphics[width=.8\columnwidth]{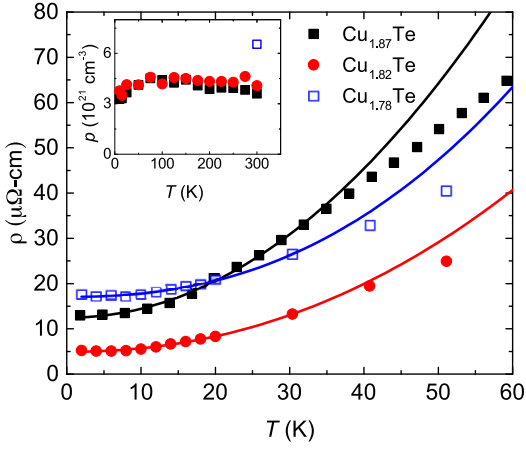}
\caption{Resistivity vs. $T$ for three Cu$_{2-x}$Te samples at low temperatures. Solid curves are $T^n$ fits as explained in text. Inset:
Hall effect-derived carrier densities.}
\label{resistivity}
\end{figure}

Figure~\ref{resistivity} exhibits resistivities below 60 K. The behavior is quadratic in the low-temperature
limit, particularly for the lowest-vacancy composition: fitting to $\rho_0+AT^n$ below 30 K yielded 
$n=2.02, 2.13$ and 2.34 for increasing $x$. Extended to room temperature, the residual
resistivity ratios for the three samples are 32, 31, and 14 for Cu$_{1.87}$Te, Cu$_{1.82}$Te, and Cu$_{1.78}$Te
respectively. The inset of Fig.~\ref{resistivity} also shows carrier densities derived from the Hall resistances, 
shown vs. temperature for Cu$_{1.87}$Te and Cu$_{1.82}$Te.
These show a low temperature downturn where the $T^2$ resistivity sets in, apparently a
result of parallel conduction paths rather than a change in the bulk carrier density, since
as noted below it is in this range that the surface states are believed to influence the transport properties.

Fixing the low-temperature exponent to $n=2$, the fitted resistivity pre-factors are $A=0.019, 0.009$, and 
0.010 $\mu\Omega$cm/K$^2$ for the three samples with increasing $x$. These compare to 
the lower end of the range for Fermi liquid behavior in heavy-Fermion materials \cite{Kadowaki1986}, 
although with the low-temperature behavior attributed to
high-mobility threading states, this implies considerably
smaller effective $A$ values for these states, comparable for example to elemental transition metals.
Mobilities derived from the resistivities and room-temperature
Hall-effect carrier densities are $\mu$ =  10 cm$^2$/Vs or less at room temperature, increasing
to 170, 280, and 55 cm$^2$/Vs for Cu$_{1.87}$Te, Cu$_{1.82}$Te, and Cu$_{1.78}$Te
respectively at low temperature. These are not unexpected for semiconductors with large vacancy densities
and large hole band mass \cite{Sirusi2016} close to 0.5 $m_e$, although the temperature dependences are large
for such a case, apparently due to threading states.

\begin{figure} \includegraphics[width=.8\columnwidth]{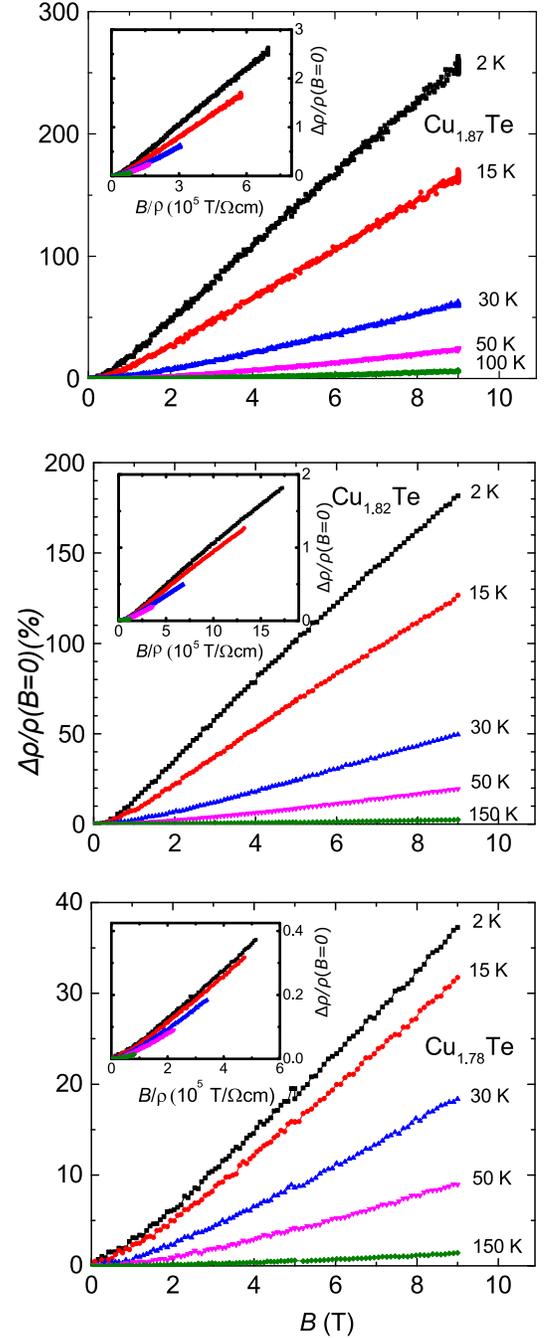}
\caption{Magnetoresistance as a function of magnetic field for the three samples at 
temperatures 2 K, 15 K, 30 K, 50 K, and 100 K or 150 K as shown. The insets are Kohler plots.}
\label{magnetoresistance}
\end{figure}

Figure \ref{magnetoresistance} displays the magnetoresistance, MR = $\Delta\rho/\rho(0)$, 
where $\Delta\rho=[\rho(B)-\rho(0)]$, and $\rho(B)$ denotes the resistivity measured in 
applied field $B$. The 2 K  
magnitudes reach 250\%, 180\%, and 37\% at 9 T for Cu$_{1.87}$Te, Cu$_{1.82}$Te, and Cu$_{1.78}$Te, respectively. 
The largest of these are comparable to effects observed \cite{Xu97} in Ag$_2$Te, although 
differing in that the Ag$_2$Te results are observed in a much narrower composition window for 
carrier densities near zero, and decrease more slowly vs. temperature.

The insets of Fig.~\ref{magnetoresistance} also display Kohler plots, 
often used to understand whether a single scattering process controls
the magnetoresistance \cite{Pippard1989}. The curves deviate from a
common line at 30 K and below, showing that there are parallel scattering processes
corresponding to the low-temperature
conduction mechanism.

\begin{figure}
\includegraphics[width=.9\columnwidth]{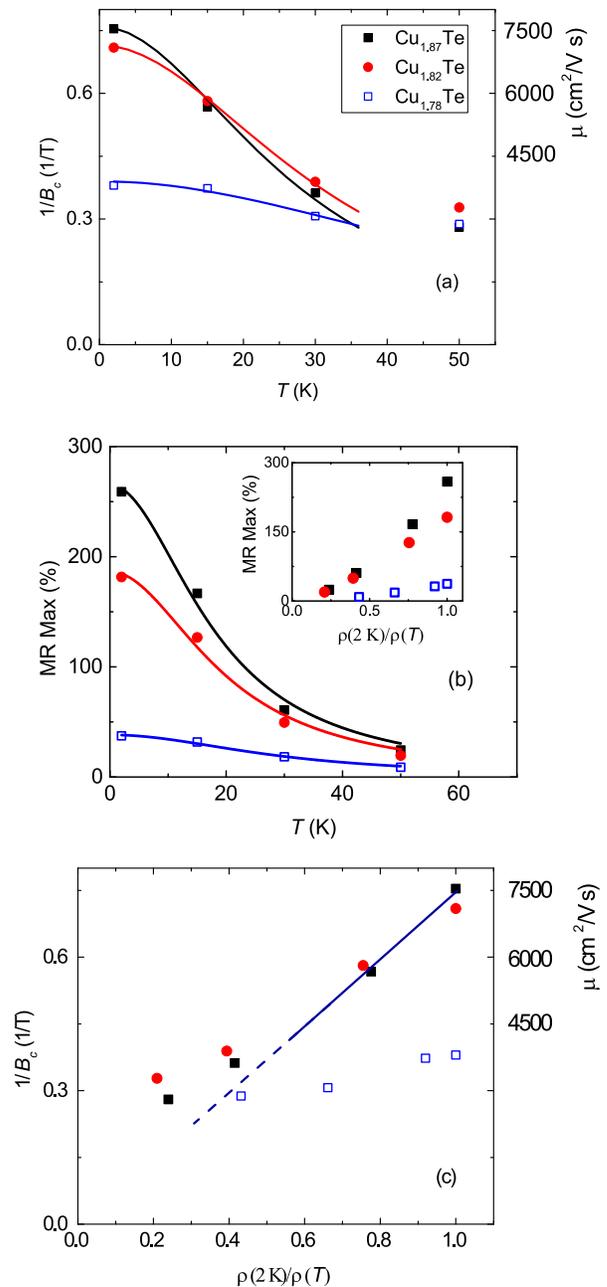}
\caption{(a) Crossover field vs. $T$ for the three samples, with mobility values corresponding to
the Parish-Littlewood crossover condition shown on the right axis. 
(b) 9 T MR values vs. $T$, and (inset) vs. scaled inverse resistivity. 
Trendlines in plots (a) and (b):  $1/(a+bT^2)$ curves. (c) Crossover field vs. normalized
inverse resistivity, with straight line through origin as guide to the eye. Symbols are common to all plots.}
\label{mobility}
\end{figure}

Fig.~\ref{mobility}(a) shows crossover fields ($B_c$), where the MR changes
from quadratic to linear. 
$B_c$ was obtained by fitting the resistivity in the low-field 
limit to a $B^2$ dependence, and linear at higher fields, and extracting the fields where these 
curves cross \cite{Narayanan2015}. The highest-temperature curves were excluded 
since the small response makes the fits unreliable. 

In the Parish and Littlewood (PL) classical transport model, $B_c$ 
corresponds to a condition $B\langle\mu\rangle = 1$.
In Fig.~\ref{mobility}(a) 
values of the average mobility are given on the right axis according to this condition. For the two lowest-$x$ 
samples these are close to 7500 cm$^2$/Vs at 2 K, considerably larger than the Hall mobilities
extracted for these samples. This confirms that there must be high mobility regions within the 
overall low mobility material, reinforcing the evidence for threading conduction due to
topological surface states, as was noted above. 

For carriers of one sign, in the PL model the magnitude of the linear MR should also scale with average mobility,
and in some cases a direct proportionality has been observed for linear MR vs. measured mobility. As a measure
of the linear MR we plotted the 9 T values (MR Max) vs. $T$ in Fig.~\ref{mobility}(b). 
In the inset these are plotted vs. the normalized inverse resistivities measured at the corresponding
temperatures, a measure of the mobilities for the case of constant carrier densities 
as we have here. However the plots are not linear, reflective of the discontinuous mobility distribution
in these samples, and the crossover in resistivity mechanisms at low temperatures.

A more direct measure of the connection between the crossover field and the high mobility carriers responsible 
for the linear MR is shown in
the plot of $1/B_c$ vs. inverse resisitivity, Fig.~\ref{mobility}(c). There is a linear relation below 30 K 
for the two lowest-$x$ samples, which exhibit the largest MR and  
low-$T$ resistivities closest to $T^2$ behavior. Thus there is strong evidence that the high
mobility threading carriers for these samples 
can be traced directly to the measured resistivity in the low temperature limit, and these carriers
apparently dominate the resistivity in this limit. The $1/(a+bT^2)$ curves in 
Fig.~\ref{mobility}(a) and (b) are also drawn to correspond to this relation, with both $1/B_c$
and MR Max connected to the inverse of the mobility for low temperatures.

\begin{figure}		
\includegraphics[width=.8\columnwidth]{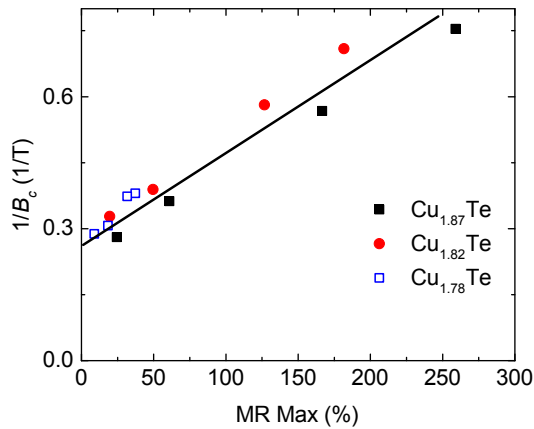}
\caption{Inverse crossover field plotted vs. maximum MR (9 T values) for the Cu$_{2-x}$Te samples at 2 K. Straight
line is guide to the eye.}
\label{figure4}
\end{figure}

A plot of $1/B_c$ vs. MR max is also given in Fig.~\ref{figure4}, and we see that there is a
universal scaling between these quantities for all samples. 
A straight-line relationship is expected for a classical transport model, but with zero intercept in the PL model. 
This result is also similar to the universal scaling identified in Ref. \onlinecite{Johnson2010}
for nanoparticle films, although again in the present case there is a large offset.
Note however that other systems have been observed to exhibit such an offset, for example in results
for Ag$_2$Se films \cite{vonkreutzbruck09} one can see that the corresponding offset has a much larger value of
about --23 T$^{-1}$. The reasons are not clear, although in the present materials the source of the
linear MR appears more likely to be misdirected currents\cite{Hu2007,kisslinger2017} rather than a broad 
distribution of mobilities, with the threading nature of the high mobility currents a distinguishing feature relative
to more continuous models. 

Aside from a surface state mechanism, 
because of its layered nature 2D weak anti-localization from spin-orbit-split bulk states could play a role in Cu$_2$Te. 
Such effects can be difficult to separate from classical linear MR,
and the corresponding magnetoconductance \cite{Hikami1980} indeed works reasonably well as an alternative 
fitting model for the present data (not shown). These fits, with addition of a large quadratic classical magnetoconductance
term, yield maximum dephasing lengths \cite{Hikami1980} at 2 K of between 19 and 30 nm, not unreasonable values. 
For comparison, in the analog material Cu$_2$Se weak anti-localization effects were also identified
at low temperatures \cite{chi14}, although with a rather different amplitude and field dependence than what is observed here. 
Thus though it seems possible that some of the observed response in Cu$_2$Te is due to
such effects, the scaling with resistivity established here points to a classical model based on 
high mobility surface states as a more reasonable model to explain these observations.

It is also possible to attribute the results to the existence of bulk Dirac-like electronic states, such as for 
the guiding center mechanism recently introduced to account for inhomogeneous transport in such
systems \cite{Song2015}. However in such cases a generally high mobility would be 
expected \cite{Liang2014,Narayanan2015,Hou2015,yang16}, as opposed to the situation here.
Alternatively, a compensation-based mechanism due to multiple carrier pockets \cite{Schnyders2015} 
might also explain the present results, however it has been shown \cite{Sirusi2017}
that hole pockets alone account well for the bulk transport and NMR behavior in Cu$_{2-x}$Te, and
the observation of linear MR in compositions with different carrier concentrations appears 
inconsistent with the balance of carrier occupations required for such a mechanism.

Another consideration would be whether magnetic quantization conditions are reached here, 
such that a quantum MR \cite{Abrikosov1998} model is appropriate. Given the effective mass obtained
for Cu$_2$Te and the carrier densities present in these samples, we expect \cite{Sirusi2016} that the Fermi
energy in the bulk corresponds to several tenths of eV. With a corresponding Fermi level for Dirac-like surface states
having the Fermi velocity of graphene, a field of 100 T or more would be required to occupy only the lowest
Landau level \cite{castroneto2009}. Thus under the conditions used here we expect that many Landau levels
will be occupied, a situation far from the quantum limit. 

In a classical treatment, it is also clear that a model based on uniformly distributed 
weak disorder \cite{Herring1960} cannot account for the 
present behavior, since the crossover fields are well out of range of the nearly-uniform high mobility 
needed for this to work. Indeed, as noted above it appears more reasonable to treat this system as
discontinuous, with carrier behavior closer to bimodal, consisting of that of the bulk and of the polycrystal 
interfaces. There have been a number of works analyzing such situations, including through
models based on resistor networks \cite{Hu2007} as well as in effective medium 
theories \cite{bergman2000,magier2006}. However we are not aware of specific predictions
related to the behavior of Fig.~\ref{figure4}, in which the behavior appears to go smoothly from
an inhomogeneous transport-based liner MR to a purely quadratic classic behavior as the
temperature and carrier density increases.

Returning to the observed $T^2$ resistivities, there has been considerable interest in understanding
the electron-electron interaction behavior of topological surface states \cite{sacksteder2014,lundgren2015}.
In some cases these are predicted to have $T^2$ behavior analogous with that of 
ordinary Fermi liquids, through processes that should be strongly dependent on umklapp scattering, and correspondingly
on the symmetry and curvature of the Fermi surface of the 2D topological states \cite{pal2012,buhrman2013}.
An alternative mechanism for the observed results would be scattering between bulk and
surface electronic states, although with a scattering rate proportional to the density of
states \cite{xu2014}, this mechanism would not be expected to produce the observed
$T^2$ dependence. However it seems possible that such a mechanism is responsible for
the much more rapid drop in linear MR with increasing bulk carrier density as observed \cite{Xu97} in 
Ag$_2$Te vs. what is seen here.
Cu$_2$Te presents a case where a clear $T^2$ behavior sets in at low temperatures dominated by the surface states,
a situation for which there are few experimental examples. 

In conclusion, we observe a large linear magnetoresistance in samples of Cu$_{2-x}$Te with increasing carrier densities,
with the magnitude reaching $\Delta\rho/\rho(0)$ = 250\% at 2 K in a 
9 T field, comparable to the effects observed in Ag$_{2}$Se and Ag$_{2}$Te. 
Examining the magnitude of the effect vs. the crossover field where the behavior changes from low-field quadratic
to high-field linear behavior, we demonstrated that models based on classical transport behavior best explain the observed
results. We also identified a universal scaling between the MR magnitude and the crossover field independent of carrier
density. The effects are traced to previously identified indications of topologically inverted behavior in this system,
such that topologically stable surface states provide the high mobility transport channels.
There is a crossover to a $T^2$ resistivity behavior at low temperatures where the large linear MR appears, 
which we connected to electron-electron interaction effects within the surface states, so this system also provides an 
experimental example of such strongly interacting surface states.

\begin{acknowledgments}
This work was supported by the Robert A. Welch Foundation, Grant
No. A-1526. Work at the University of Michigan were supported as
part of the Center for Solar and Thermal Energy Conversion, an Energy Frontier Research Center funded by
the U.S. Department of Energy, Office of Basic Energy Sciences under Award DE-SC-0000957.
\end{acknowledgments}

\bibliography{Cu2Te}

\end{document}